# Characterization of Instrumental Phase Stability

D. Y. Kubo[*a], T. R. Hunter[b], R. D. Christensen[c], P. I. Yamaguchi[c]

[a]Academia Sinica, Institute for Astronomy & Astrophysics, 645 N. Aohoku Pl., Hilo, HI, USA
[b]Harvard-Smithsonian Center for Astrophysics, 60 Garden St., MS 78, Cambridge, MA, USA
[c]Smithsonian Submillimeter Array, 645 N. Aohoku Pl., Hilo, HI, USA 96720-2700

## ABSTRACT

Atmospheric water vapor causes significant undesired phase fluctuations for the SMA interferometer, particularly in its highest frequency observing band of 690 GHz. One proposed solution to this atmospheric effect is to observe simultaneously at two separate frequency bands of 230 and 690 GHz. Although the phase fluctuations have a smaller magnitude at the lower frequency, they can be measured more accurately and on shorter timescales due to the greater sensitivity of the array to celestial point source calibrators at this frequency. In theory, we can measure the atmospheric phase fluctuations in the 230 GHz band, scale them appropriately with frequency, and apply them to the data in 690 band during the post-observation calibration process. The ultimate limit to this atmospheric phase calibration scheme will be set by the instrumental phase stability of the IF and LO systems. We describe the methodology and initial results of the phase stability characterization of the IF and LO systems.

**Keywords:** phase stability, phase transfer, local oscillator, YIG oscillator, Gunn oscillator

## 1. INTRODUCTION

The Submillimeter Array (SMA) is a collaborative project of the Smithsonian Astrophysical Observatory (SAO) and the Academia Sinica Institute of Astronomy & Astrophysics (ASIAA) of Taiwan. The array consists of eight six-meter diameter antennas with receivers operating from 180 to 700 GHz and a digital correlator with 2 GHz of bandwidth. Located on Mauna Kea, Hawaii, the primary elements of the SMA interferometer can be reconfigured across twenty four pads which provide baselines ranging from 8 to 508 meters. Each antenna cryostat assembly houses up to eight receiver inserts consisting of low-noise superconducting (SIS) mixers[1]. The inserts can be used in pairs for increased bandwidth and for polarimetry observations. Local oscillators (LOs) are derived in each antenna to provide heterodyne mixing from the sky frequency to the 5 GHz intermediate frequency (IF). In this paper, we describe the IF and LO systems and their associated instrumental phase stability.

## 2. IF/LO SYSTEM DESCRIPTION

The IF/LO hardware is physically partitioned between the main control building and the antenna cabins. All signals transmitted to and from the antennas are through fiber optic cables. Each of these areas is described in the following subparagraphs.

### Control Building IF/LO System

The control building IF/LO electronics is located in a shielded room to reduce EMI effects. The analog equipment is mounted in thirteen racks and is physically separated from digital correlator. Figure 1 shows a photograph of a portion of these analog racks. Cooling and thermal regulation of the racks is provided by an air conditioning system dedicated to this purpose. Each rack has independently servo controlled louvers to maintain temperature regulation. A functional block diagram of the analog room equipment is provided in Figure 2. Beginning with the LO system, a 10 MHz crystal oscillator is phase locked to a GPS reference to maintain a high level of frequency accuracy and is the master reference generator (MRG) for the entire LO system. One set of ~109 MHz and 200 MHz references are generated then optically modulated and sent out to each of the eight antennas via fiber-A for the 200/300 GHz band receiver LO. A similar set is generated and sent out via fiber-B for the 400/600 GHz band receiver LO. The ~109 MHz reference is generated by direct digital synthesis (DDS) to produce unique Walsh functions[2] and primary fringe stopping. There are sixteen uniquely generated ~109 MHz references for the eight antennas. The 200 MHz references are common to all. A reference LO consisting of a tunable YIG oscillator, denoted as MRG YIG-1 PLL in Figure 2, is provided to each of the eight antennas through an optical 10-way power divider. The two additional fiber-C outputs are provided to the CSO and JCMT for extended SMA operations. MRG YIG-1 is used as a reference for the 200/300 GHz band receiver LO and MRG YIG-2 for the 400/600 GHz band receiver LO. Both YIGs are tunable from 5.5 to 8.5 GHz but since they reside on the same fiber-C path they must be tuned to maintain a minimum typical frequency separation of 200 MHz.

*dkubo@asiaa.sinica.edu.tw; phone 808-961-2926; fax 808-961-2989; www.asiaa.sinica.edu.tw

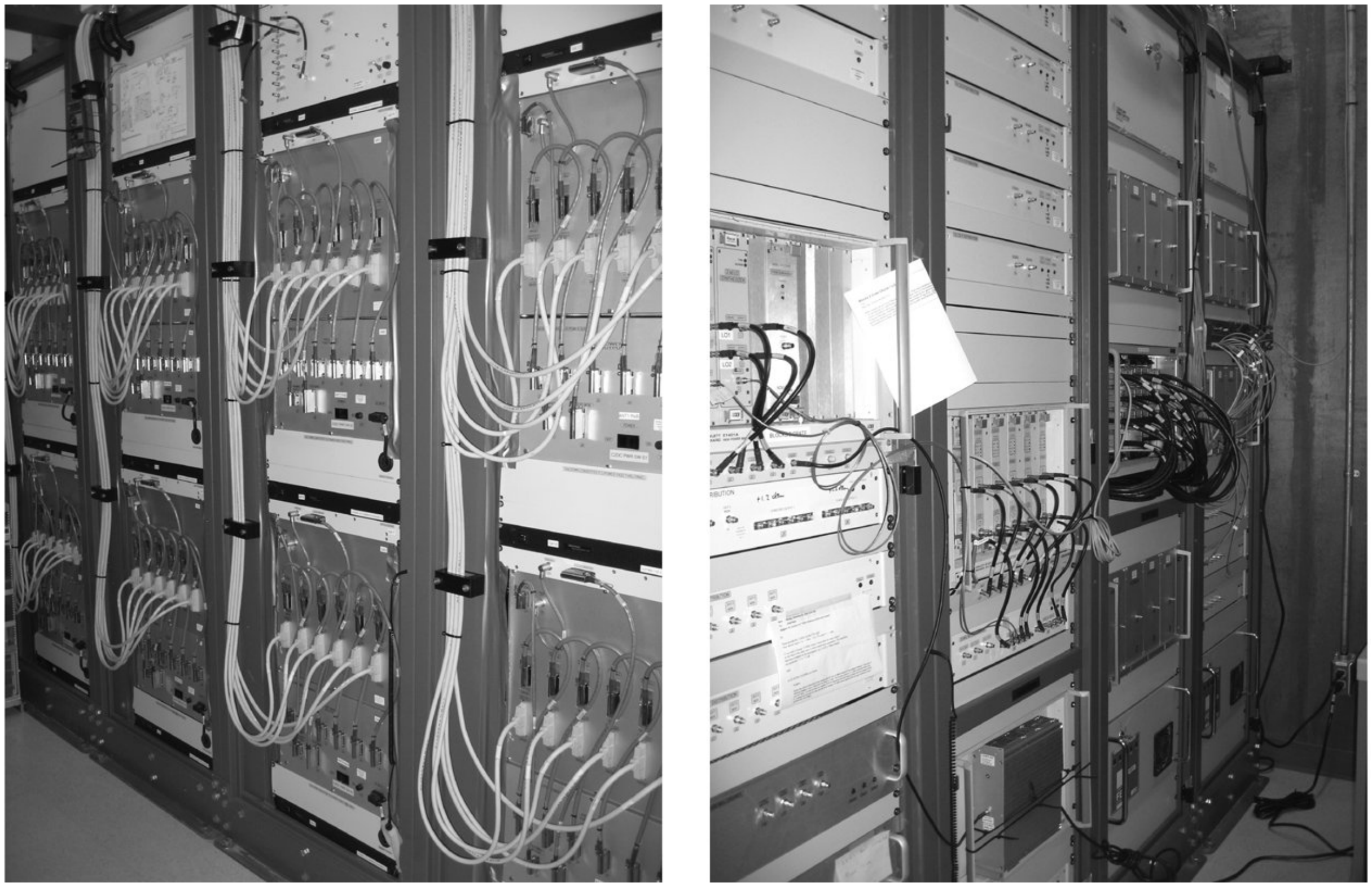

Figure 1. Left photo – rear view of analog racks 1 through 4 consisting of IF-1 first and second down converters for antennas 1 through 8. Right photo – front view of analog racks 10 through 13 consisting of the LO reference system. The racks are thermally regulated to maintain constant equipment temperature.

Fibers A and B, in addition to carrying the ~109/200 MHz LOs to the antennas, also provide the return path for the IF signals. This is accomplished by using 1550 nm optical transceivers for the ~109/200 MHz signals, and 1310 nm transceivers for the IF. Wave division multiplexers (WDMs) are used to separate the wavelengths at each end. The 5 GHz IF signal is processed through the first down converter which provides signal leveling followed by down conversion into six block outputs centered at 1 GHz (328 MHz bandwidth). The signal leveling is accomplished by a variable attenuator which is set only at the beginning of an astronomical track using a "setIFLevels" software routine. The down conversion is accomplished by using six mixers with separate first LOs of: 3180; 3508; 3836; 6164; 6492, and 6820 MHz. The second down converter further subdivides each block into four chunks centered at 153 MHz (82 MHz bandwidth) using four mixers with second LOs of: 724; 806; 1194; and 1276 MHz. Twenty four separate complex phase modulators are applied to the chunk LOs for secondary fringe stopping. Each of the twenty four 153 MHz IF signals are leveled again using variable attenuators to maintain optimal drive levels to the analog-to-digital converters (ADC) which follow. The ADCs provide 2-bit sampling at 208 Msps, and as a result of the coarse sampling have a fairly narrow input dynamic range. The chunk leveling loop can be software configured to provide continuous update to accommodate for varying sky conditions. Following the ADCs, the signals are demultiplexed-by-four and sent to the digital correlator as differential ECL (emitter coupled logic) through differential twisted shielded pair cables. The left photo of Figure 1 shows these white sampler cables, each representing four chunks of data.

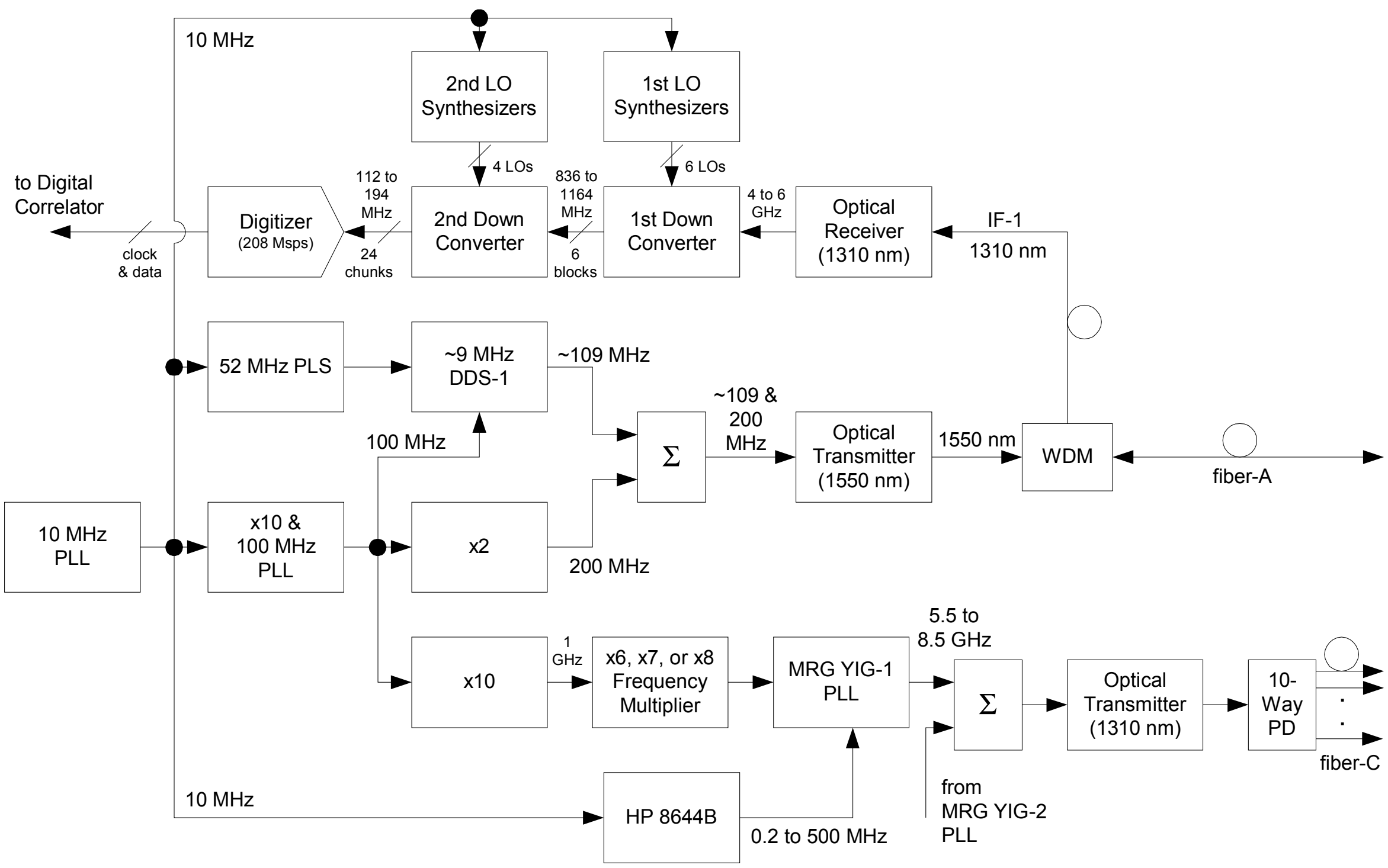

Figure 2. Block diagram of IF/LO-1 system located in analog room. IF-1 is received through fiber-A then down converted and separated into 24 chunks centered at 153 MHz. Each 82 MHz wide chunk is digitized and sent to the digital correlator. The ~109/200 MHz LO references are generated and sent to the antennas through fiber-A. A separate set of ~109 MHz signals are generated for each antenna. A tunable 5.5-8.5 GHz LO-1 is generated and optically divided and sent to each antenna through fiber-C. IF/LO-2 requires a similar set of hardware.

## Antenna IF/LO System

The antenna cabin IF/LO electronics are housed within an enclosure mounted to the cabin wall. Figure 3 shows photographs of this enclosure. Cooling and thermal regulation is provided through passive means using the cabin air environment. The cabin environment itself is stringently regulated using an air handler system which maintains the temperature to within +/-0.1 degrees C RMS over a period of several hours. A functional block diagram of the cabin electronics is provided in Figure 4. Each antenna receives a separate set of ~109/200 MHz LOs through fibers A and B, along with a pair of tunable 5.5 to 8.5 GHz LOs on fiber-C. The YIG-1 PLL phase locks to the 200 MHz and LO-1 and produces an output at either LO-1 +/- 200 MHz. The same "setIFLevels" command mentioned earlier sets attenuators within the ~109/200 MHz and MRG LO-1 paths to provide optimal drive levels to the YIG PLL. This optimal drive level ensures repeatable low phase noise performance of the YIG PLL, however, it is important to keep the attenuators at a fixed value during an astronomical track for phase stability. A typical example frequency for a 230 GHz observation would be LO-1 at 6.825 GHz and the YIG-1 locked to 7.025 GHz. The antenna YIG is not permitted to operate below 6 GHz to prevent in-band leakage into the IF system. Following the YIG-1 output is a harmonic mixer which produces multiple harmonics (M) and mixes them with the Gunn oscillator output operating in the 80 to 120 GHz band. M for the 230 GHz example is 16 which produces 112.40 GHz. Through the action of the Gunn digital PLL[3] (which is usually operated in USB) the Gunn oscillator will be locked at M*YIG1 + 109 MHz, which is 112.509 GHz for the 230 GHz example. The Gunn output is followed by a final fixed multiplier whose value is denoted as N. N for this example is 2 making the final LO to the receiver 225.018 GHz. A photo of the Gunn PLL assembly is shown in Figure 5. The Gunn PLL assemblies are located along the top of the optics cage and are held at a very constant operating temperature due to the performance of the air handler system. The center of the IF-1 band will correspond to an USB sky frequency of 230.018 GHz. Table 1 provides example LO frequencies for some typical tuning frequencies. It becomes obvious from the table that a small movement in the antenna YIG phase translates to a non-trivial phase movement at the final LO. For example, a 1.0 degree phase movement at the antenna YIG output translates to 32, 48, and 84 degrees at the final LO for 230, 345, and 690 GHz, respectively, which is clearly too large because it will correspond to significant errors in the position measurement of astronomical sources and to a reduction of the image fidelity.

The 5 GHz IF signal from the low frequency receiver is processed through the IF-1 channel shown in Figure 4. The IF signal is leveled using a variable attenuator to provide optimal drive to the fiber optic transmitter and can accommodate for receiver output power variations as a result of specific tuning conditions. This variable attenuator is set only once at the beginning of a track using the "setIFLevels" command. A duplicate set of hardware is provided for IF-2 for processing the high frequency receiver IF. Unlike the LO system, the phase of the IF system does not get scaled thus we are not very concerned about it as long as it remains small. Our test methodology for characterizing the LO system, however, does rely on the IF system to behave in a manner suitable for detecting returned LO phase changes of a few tenths of a degree over several hours. Test results described later indeed confirms this to be the case.

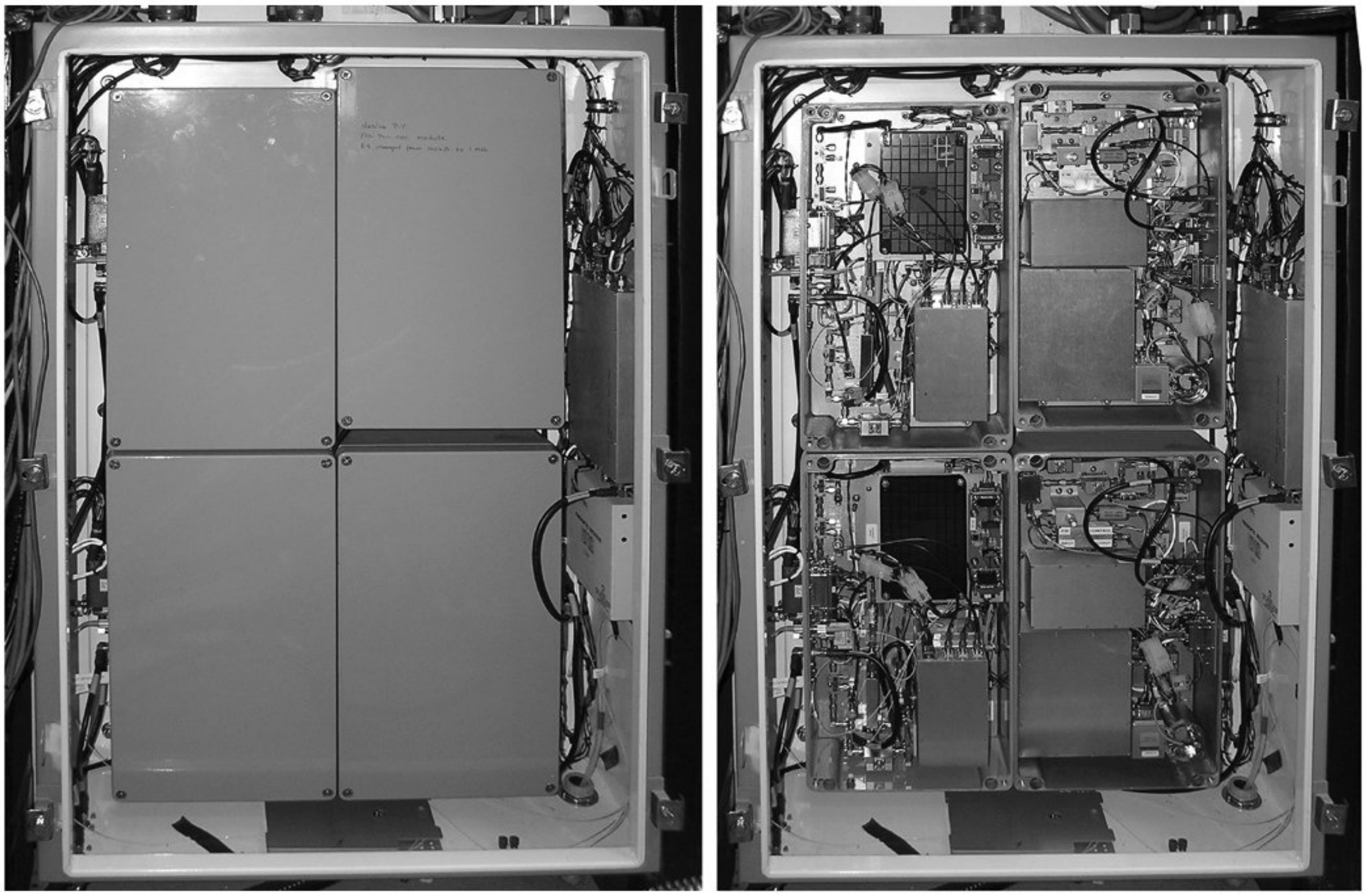

Figure 3. Left photo – antenna cabin IF/LO electronics cabinet (36 x 24 x 9 inches) with outer cover removed. Right photo – internal covers removed. The upper two boxes are IF-1 (left) and YIG-1 (right), lower two boxes are IF-2 and YIG-2. The assembly on the right side wall of the enclosure is the LO receiver plate. The three single-mode fibers enter through the fitting visible on the bottom right.

**Fiber System**

There are a total of three cables provided to each antenna from the control building which consist of a single-mode fiber cable (3 fibers), multi-mode fiber cable (12 fibers), and 480 VAC power cable (4 wires). Since the eight antennas can reside on any of the twenty four pads, fiber interconnections are made at both the pad and a patch panel within a fiber optic vault located in the basement of the control building. The multi-mode fibers are used for a variety of functions including the control of the electronics and mechanical servos as well as the return of telemetry information.

The single-mode fibers are used exclusively by the IF/LO system. These fibers have been manufactured by Sumitomo to have a low transmission delay coefficient with respect to temperature. The specification is </= 3.3 psec/(km*degree C). The cable runs to each of the 24 pads are buried in conduit at a depth of approximately one meter with the longest length being 734 meters to pad 20. Diurnal fluctuations in the middle of the long conduit runs have been measured to be less than 0.01 degrees C RMS[4]. Using this number as an example, a 0.01 degree C change in temperature along the entire 734 meter length of fiber will result in 0.07 degrees RMS of phase movement at an LO frequency of 7.5 GHz. A critical area for the single mode-fibers is the transition from the fixed pad base to the rotating antenna cabin. It was determined that phase stability through these Sumitomo fibers were much less sensitive to pure torsional motion than to bending or elongation. Thus the special fiber tension assembly[5] shown in Figure 6 was developed for the SMA.

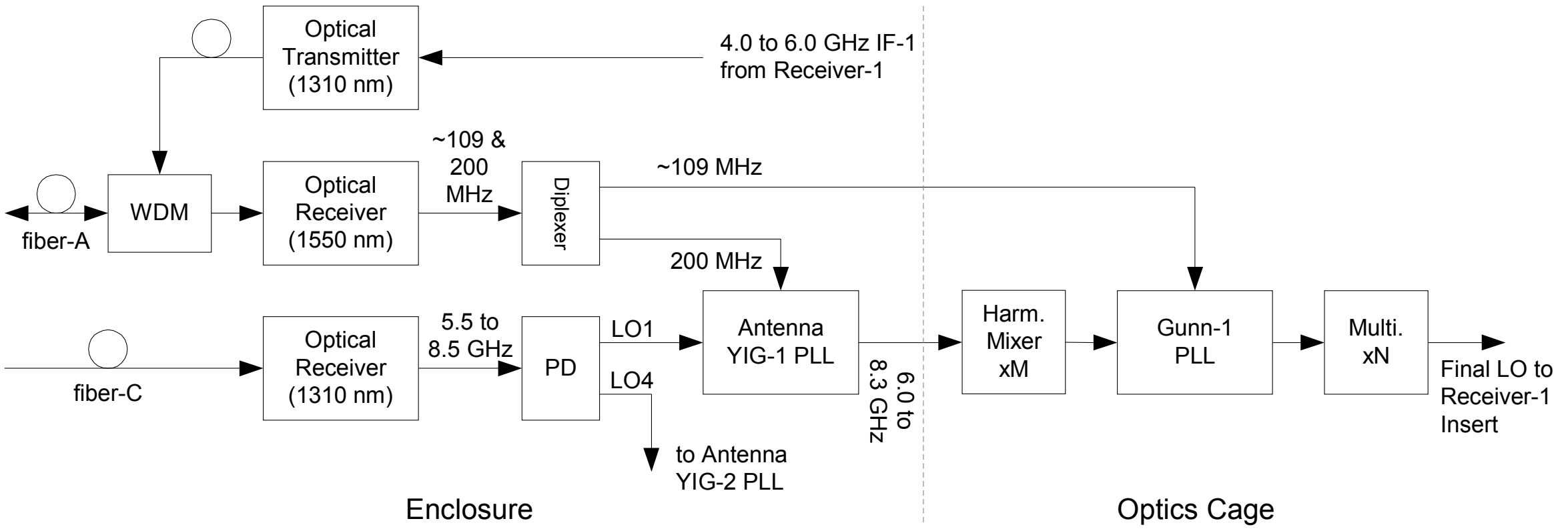

Figure 4. Block diagram of IF/LO-1 system located in antenna cabin. The harmonic mixer, Gunn PLL, and multiplier are physically located on the top of the optics cage. IF/LO-2 requires a similar set of hardware.

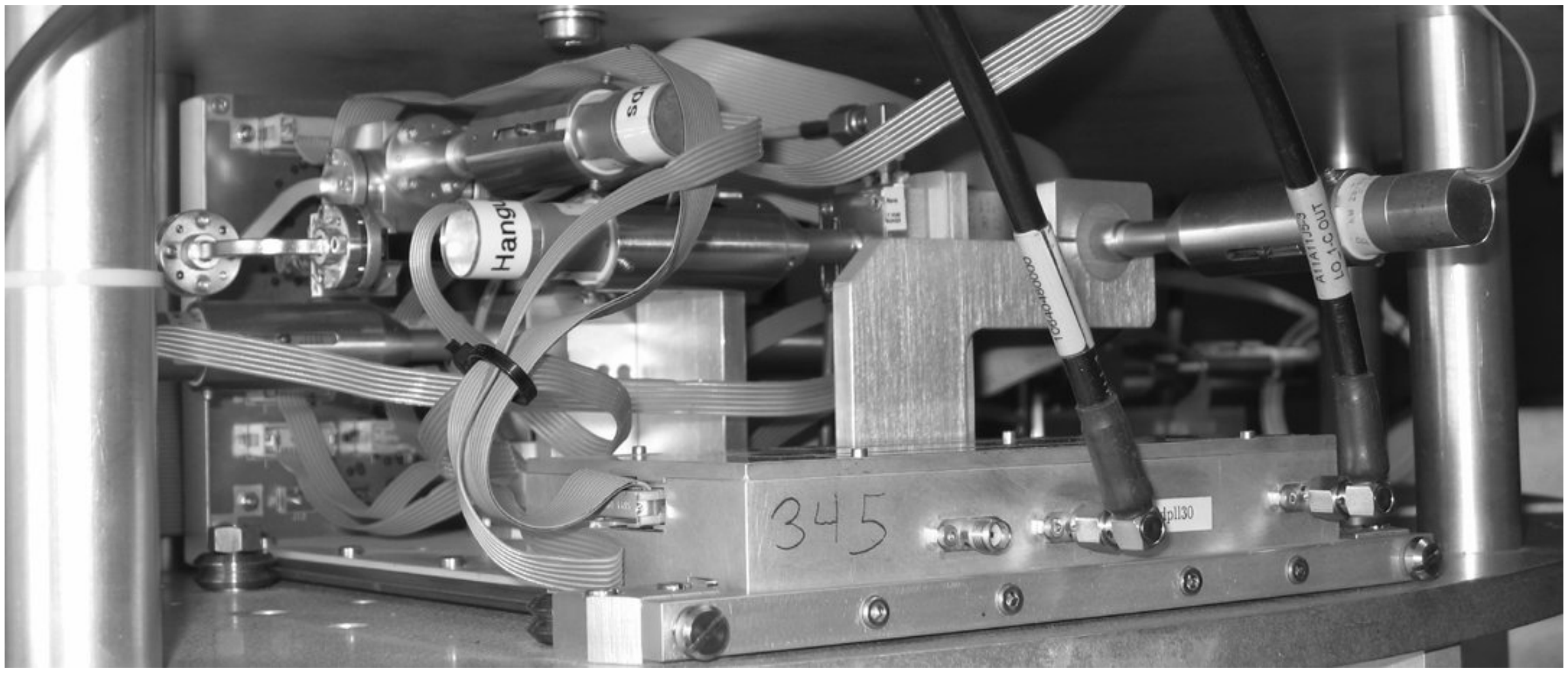

Figure 5. Photo of the antenna 300 GHz LO chain including a motorized Gunn oscillator, digital PLL, and microcontroller. Located in each antenna, there is one LO chain per frequency band.

Table 1. Example LO frequencies for 230, 345, and 690 GHz.

| MRG YIG | Antenna YIG | M | Gunn | N | Final LO |
|---|---|---|---|---|---|
| 6.825 GHz | 7.025 GHz | 16 | 112.509 GHz | 2 | 225.018 GHz |
| 6.877 GHz | 7.077 GHz | 16 | 113.341 GHz | 3 | 340.023 GHz |
| 7.947 GHz | 8.147 GHz | 14 | 114.167 GHz | 6 | 685.002 GHz |

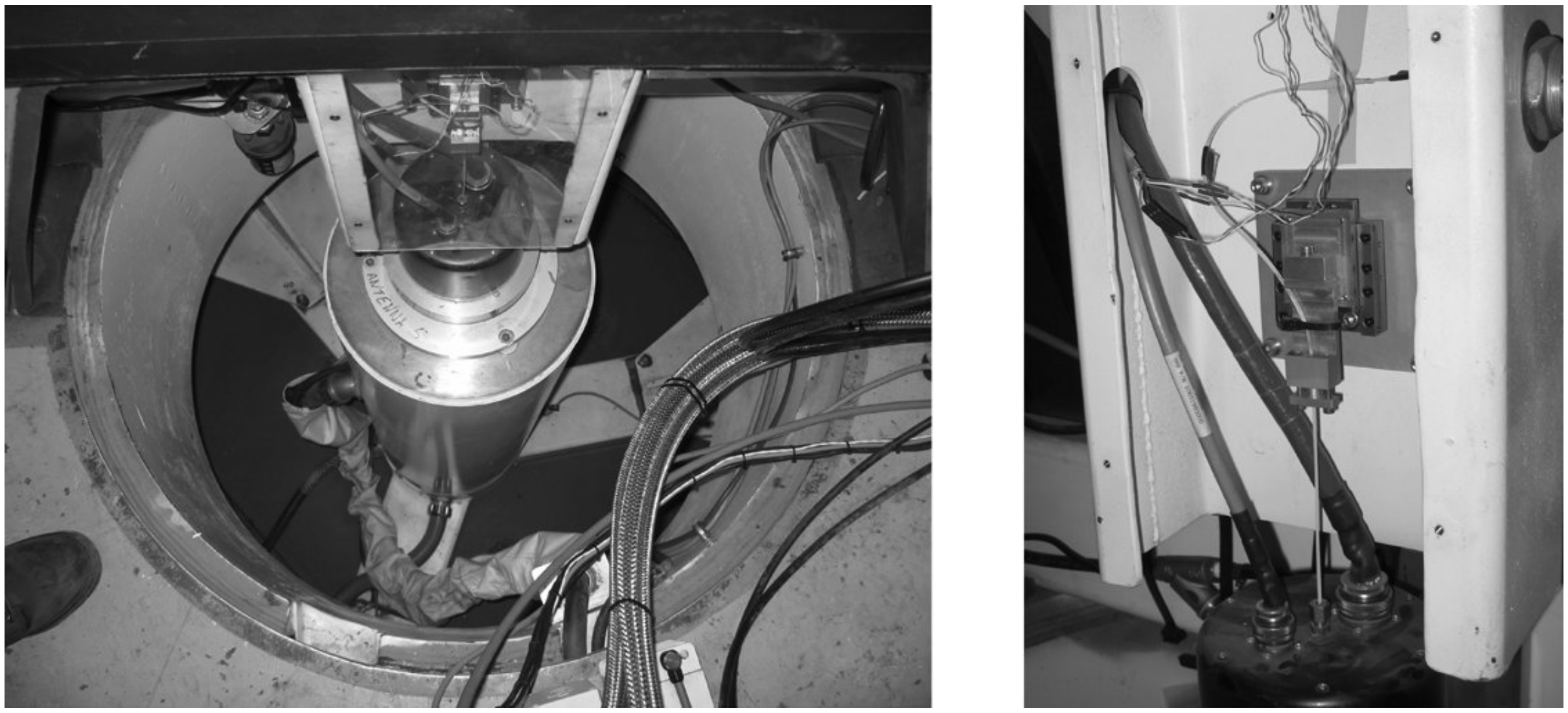

Figure 6. Photographs of the azimuth fiber tension assembly. Left – azimuth encoder (black cylindrical assembly) shown just below tension assembly. Right – tension assembly with three fibers shown coming up through a hole located in center of the azimuth encoder. The fibers are fixed at the bottom of the azimuth column and a constant tension is applied to the fibers through a spring loaded mechanism.

## 3. PHASE STABILITY PERFORMANCE

**Baseline Phase Calibration**

The primary instrument used to measure phase stability is the HP 8508A vector voltmeter which accepts two inputs from 300 kHz to 2 GHz. This vector voltmeter is installed within AR9 (analog rack 9) for temperature stability. The instrument compares amplitude and phase between a reference input and the signal under test. Readout of the data is provided through GPIB using a program which logs the information at a rate of 1 sample/second. An initial test was conducted to characterize the stability of the measurement system and associated interconnection cables. For this test the 200 MHz MRG signal was connected to port-A and the 200 MHz output for antenna 5 DDS-1 to port-B of the vector voltmeter. A plot of the phase stability is given in Figure 7 and shows a phase movement of 0.05 degrees peak to peak over a duration of three hours. The calculated standard deviation or RMS value over this same interval was 0.005 degrees. The keys to achieving this stability were to install the vector voltmeter within the temperature stabilized analog rack and to close the doors to the analog room so that the interconnect cables were exposed to a fairly constant room temperature. It took several hours for the analog room to stabilize in temperature after the door was closed.

**Antenna YIG-1 Phase Stability**

This test was developed to characterize the absolute phase stability of the antenna YIG-1 LO. It was accomplished by temporarily coupling the YIG-1 LO output in antenna 4 and sending it back to the analog room via IF-2, which was otherwise not in use at the time. Some minor hardware modifications were necessary to permit the higher LO frequencies to pass through the return IF which normally processes 4 to 6 GHz. A block diagram for this setup is provided in Figure 8. This test captures phase movements in the path which is external to the MRG YIG-1 function. I.e., it includes effects from: the MRG FOTx (MRG fiber optic transmitter); fiber-A including effects of the azimuth tension assembly; the fiber optic receiver; and the antenna YIG-1 along with the 200 MHz reference derived from fiber-A. As an undesired byproduct it also captures the phase of the return path which includes: the antenna IF-2 assembly; fiber-B and the azimuth tension assembly; the C1DC-2 (correlator 1$^{st}$ down converter); and the external mixer used to mix MRG YIG-1 LO with the returned antenna YIG-1 LO. What this test does not capture is any movement of the MRG YIG-1 LO.

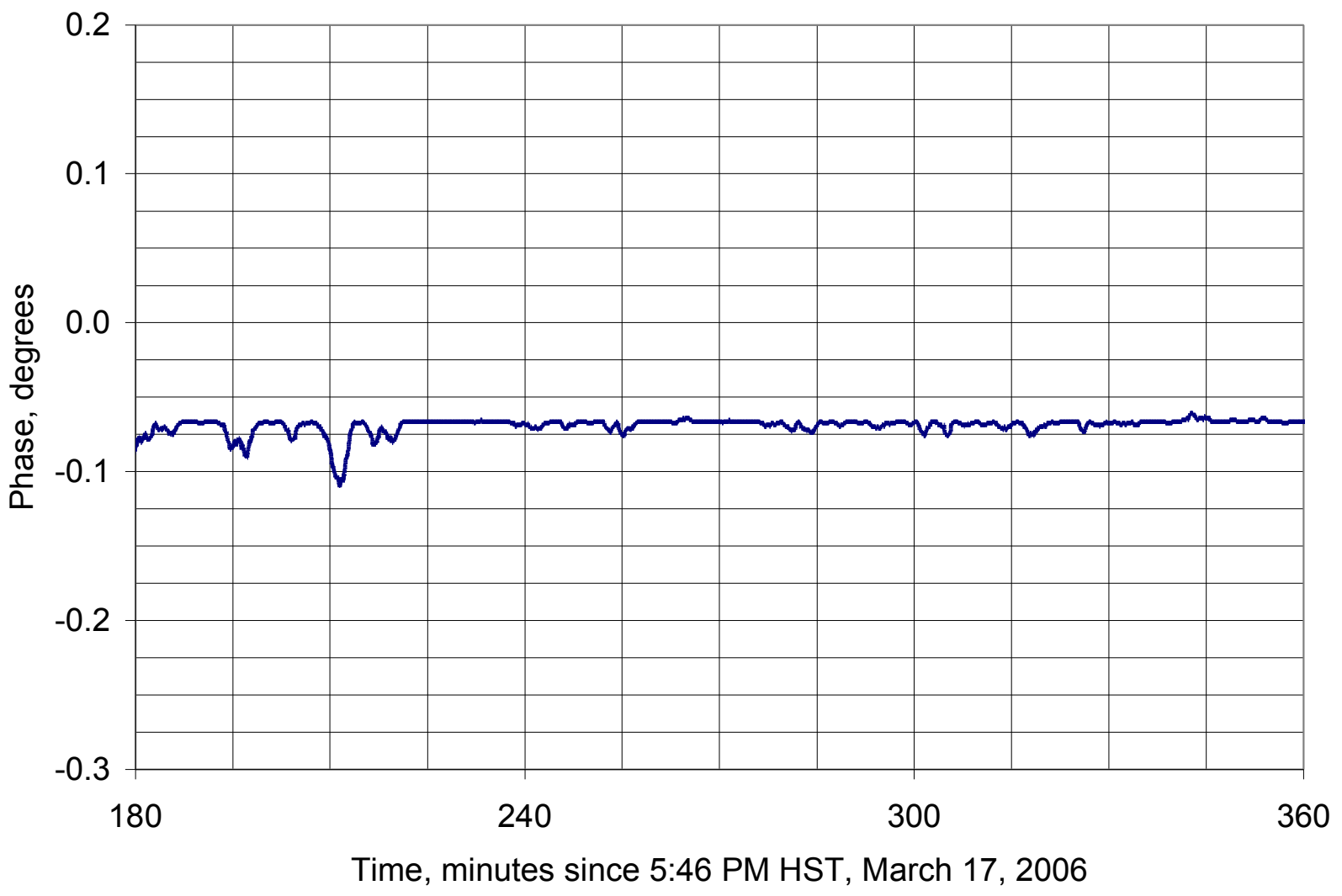

Figure 7. Phase stability of plot of baseline test over a duration of three hours. Peak to peak phase variation was measured to be 0.05 degrees over this period with a calculated RMS value of 0.005 degrees.

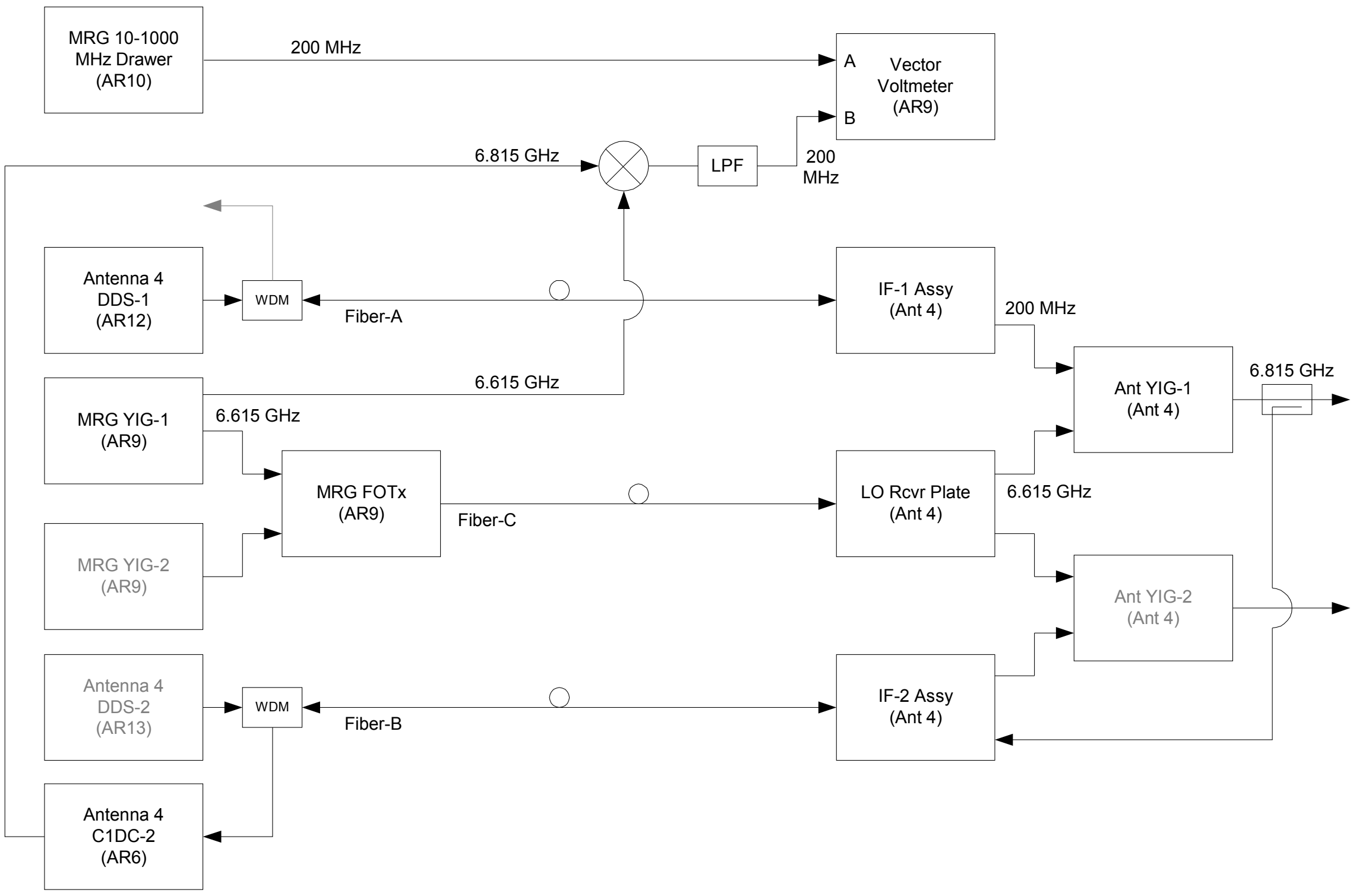

Figure 8. Antenna YIG-1 phase stability test setup. Antenna YIG-1 LO in antenna 4 was connected to IF-2 for return to analog room via fiber-B. Received LO is mixed with MRG YIG-1 LO to generate a difference frequency of 200 MHz and phase compared to the MRG 200 MHz reference.

Phase stability tests were performed on antenna 4 during an actual astronomical track. The left and right plots of Figure 9 represent the phase and antenna azimuth position, respectively. The round trip phase movement over a duration of three hours was measured to be 0.45 degrees peak to peak, with a calculated value of 0.101 degrees RMS. Some periodicity is seen which is believed to be associated to analog room and/or rack temperature variations in AR6 (houses 1st down converter for return path). We presume that the actual output phase stability of the antenna YIG-1 LO may be better than this number because it includes the effects of the return IF path. Assuming that both outgoing and return paths contribute equal amounts of phase instability, the one way value should be 0.071 degrees RMS. One useful piece of information from this test is that the return IF path phase drift must be less than 0.45 degrees which is small enough to ignore. The phase plot shows no obvious relation to the azimuth position and implies that the fiber tension assembly is working well. An identical test was conducted on antenna 5 with much different results shown in Figure 10. A relatively large phase movement of 2.30 degrees is seen over the three hour duration and shows a strong relationship of approximately 4 degrees electrical phase per 100 degrees of azimuth rotation. Initial investigation of the antenna 5 azimuth fibers revealed a loose tension assembly as well as some kinked fibers. The poor results from antenna 5 provides motivation to conduct this test for all antennas to confirm proper functionality of the LO reference hardware.

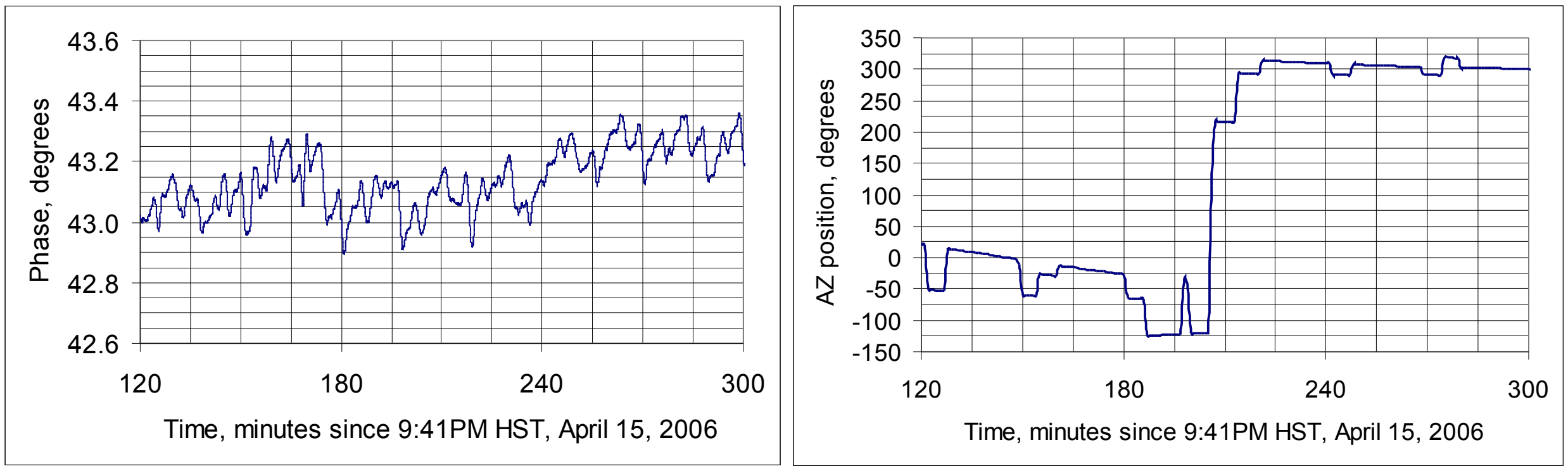

Figure 9. Left - phase stability of plot of antenna 4 YIG-1 LO after returned back to analog room. Round trip peak to peak phase variation over three hours was measured to be 0.45 degrees with a calculated value of 0.101 degrees RMS. This data was taken during an actual astronomical track. The periodic variations over ~15 minutes are believed to be associated to temperature variations within the analog room. Right – corresponding azimuth position for same antenna.

**MRG YIG-1 Verses MRG YIG-2 LO Phase Stability**

Since the MRG YIG-1 LO is derived in the analog room and is optically divided ten ways to each antenna, phase movement of this LO affects each antenna equally. The same holds true for the MRG YIG-2 LO. However, the phase drift between these two references is also of concern. Due to increased atmospheric absorption and less sensitive receivers, calibration of the interferometer data becomes more difficult at higher frequencies. With the dual receiver/IF system of the SMA, the possibility exists to transfer the interferometer calibration from the low frequency band to the high frequency band using an atmospheric model. In order for this method to work, the phase drift between the two reference signals must be kept very small. A test setup was configured to compare the phase stability between the two MRG YIG LOs by tuning both to the same frequency of 7.1 GHz. These two LOs were mixed with the 7.0 GHz output of the x7 frequency multiplier (refer to Figure 2) to down convert to 100 MHz for phase comparison using the vector voltmeter. Since this test could not be performed during an astronomical track it was run during the day with results shown in Figure 11. The rack air conditioning system turned on at approximately 8:55 AM HST (Hawaii Standard Time) in response to warmer conditions during the day and has affected the data. This data, nevertheless, shows a peak to peak variation over the first 45 minutes of 0.28 degrees and an RMS of 0.058 degrees. The data over the 45 to 180 minute range shows 0.90 degrees peak to peak with an RMS of 0.222 degrees. Assuming equal contributions of phase instabilities for MRG LO-1 and MRG LO-2, we can assign a value of 0.041 degrees RMS to each over the first 45 minutes.

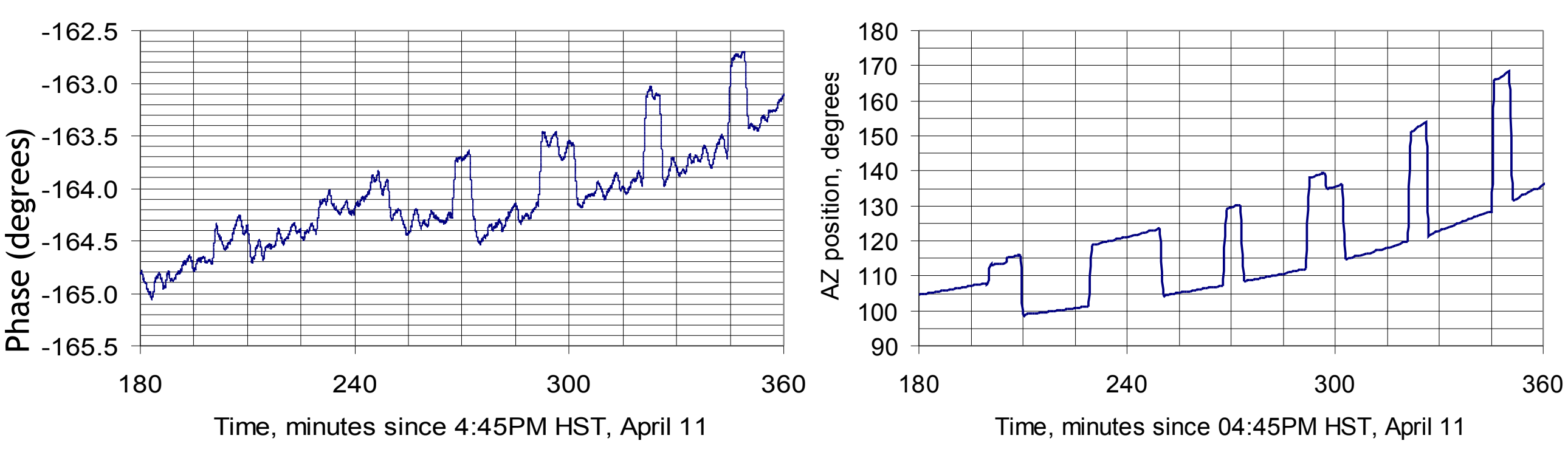

Figure 10. Left - phase stability of plot of antenna 5 YIG-1 LO during an actual astronomical track. Note the large phase change of 2.30 degrees peak to peak over the three hour duration. The calculated RMS value was 0.486 degrees. This variation is associated to the azimuth position and was traced to a loose fiber tension assembly in this antenna. Right – corresponding azimuth position.

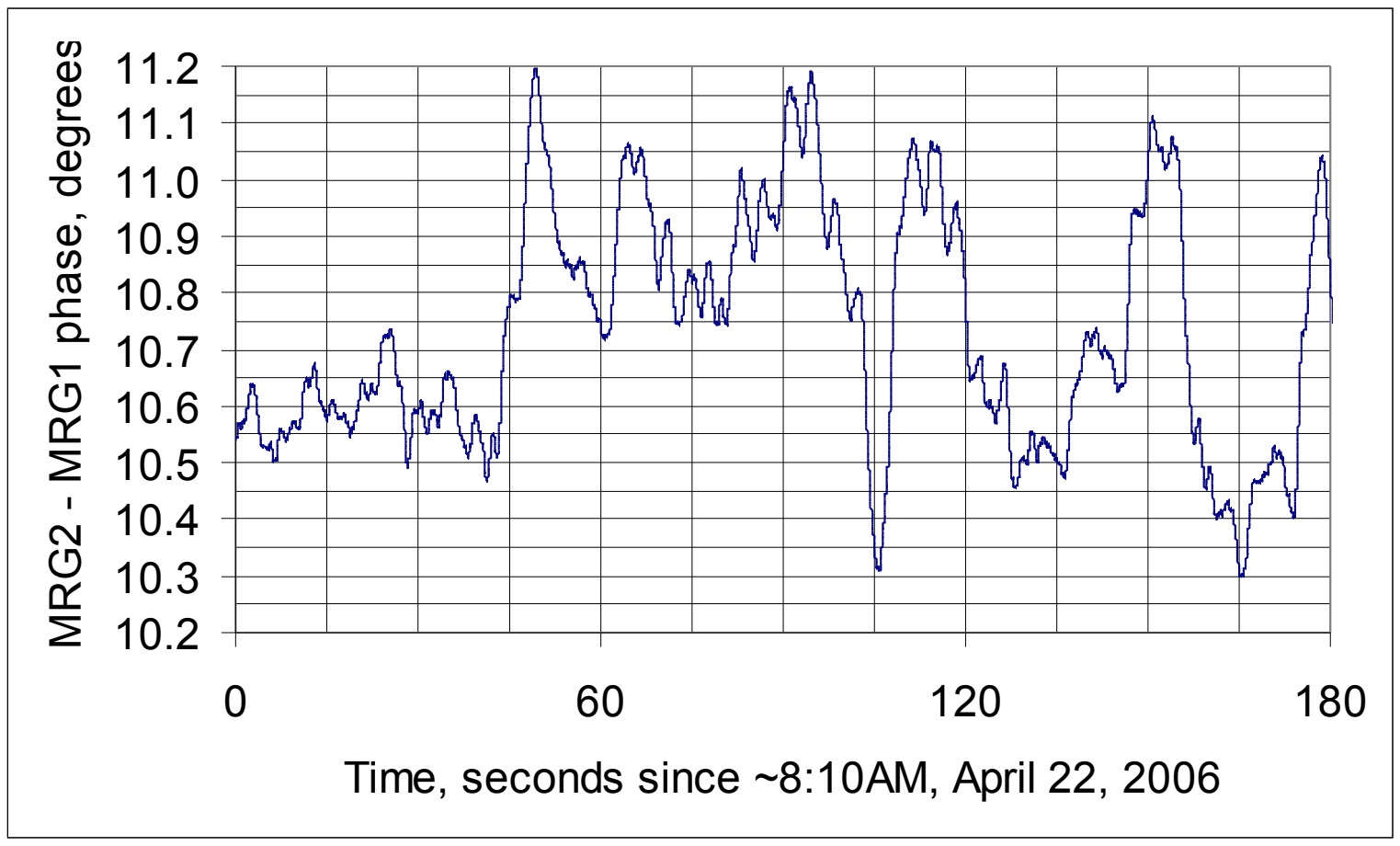

Figure 11. Phase difference between MRG LO-2 and MRG LO-1. This test was conducted during the day and as a result was subject to cycling of the air conditioning compressor which began at approximately 45 minutes after the beginning of this test run. The calculated RMS values for 0 to 45 minutes and 45 to 180 minutes are 0.058 and 0.222 degrees, respectively.

## YIG Oscillator Phase Verses Temperature

A large number of components used within the LO system show a phase change with respect to temperature. Thermal stabilization of the equipment is provided at both locations, however, some improvement is still in work. The air conditioning servo control for some of the analog racks requires further refinement to minimize temperature variations. Within the antenna cabin the IF/LO enclosure has been found to be coupled to the outside temperature through the cabin wall. This has been mostly rectified in antenna 6 with the use of close cell foam insulation between the IF/LO enclosure and the cabin wall.

One issue which affects the phase stability of the LOs at both locations is that the power dissipation of the Micro Lambda YIG oscillator varies as a function of the tuning frequency. A higher tuning frequency requires a higher coil voltage and results in a warmer operating temperature. Figure 12 provides plots of the YIG temperature (left) and the control voltage (right). Note the increase in the YIG temperate at 90 minutes when the control voltage was increased a modest amount from 3.8 to 5.5 V (total control range is 0 to 10 V). It is also evident that even though the control voltage is held constant from 360 minutes the YIG temperature continues to fall by another 0.25 C over the next six hours because it is still weakly coupled to the outside air temperature. Oven tests of the YIG assembly have shown that the output phase increases by approximately 0.12 degrees per 1.0 degree C increase in temperature. We have since discovered that increasing the PLL low frequency gain by a factor of ten has reduced the output phase variation to approximately 0.02 degrees per 1.0 degree C increase in temperature. Since temperature variations of the YIG oscillator itself affect other components within the enclosure, we are currently developing a circuit which provides constant power dissipation for the YIG which is independent of the control voltage.

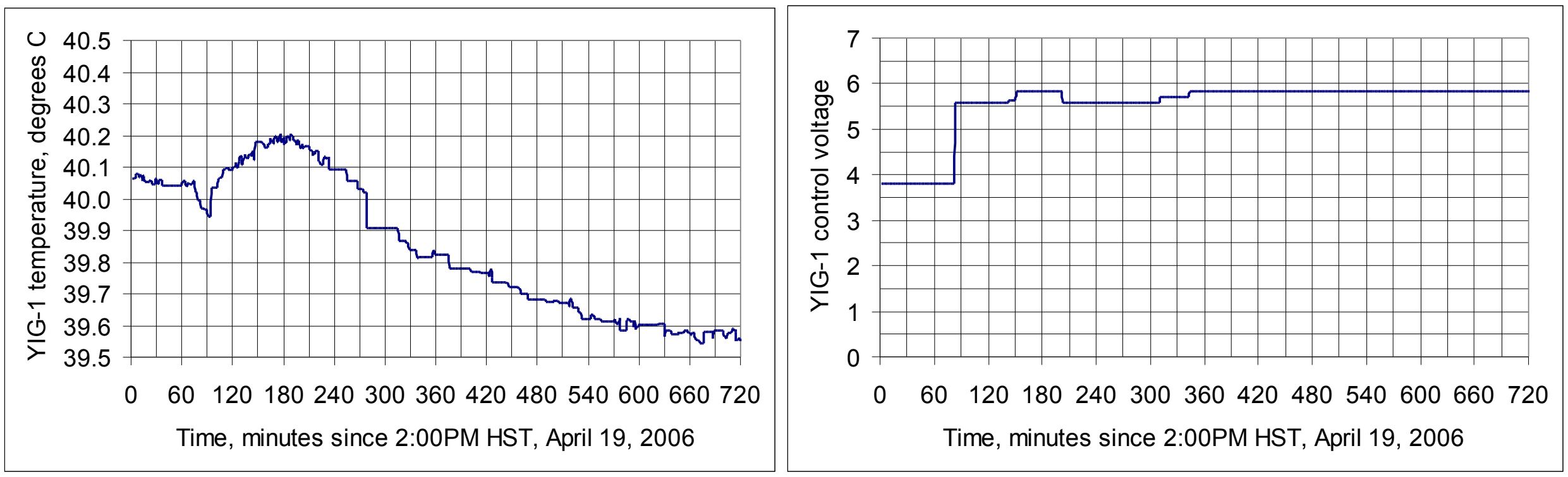

Figure 12. Left - antenna 6 YIG-1 temperature over a 12 hour interval. Right – antenna 6 control voltage over same interval.

## 4. PRELIMINARY CONCLUSIONS

The design of the SMA LO system utilizes references within the antenna which are multiplied up in frequency by 32 and 84 for observations at 230 and 690 GHz, respectively. A phase measurement for the antenna 4 LO-1 reference has shown a stability of 0.071 degrees RMS over a duration of three hours. In addition, a phase stability of 0.041 degrees RMS was obtained for MRG LO-1 over a limited 45 minute interval. Combining these two values in root sum squared fashion produces 0.082 degrees RMS for the LO-1 reference prior to multiplication. This translates to a final phase stability of 2.62 degrees RMS for the 230 GHz LO. Assuming that the LO-2 reference has similar performance, we should see a final phase stability of 6.89 degrees RMS for the 690 GHz LO. These numbers should be acceptable for transferring the phase observed from the 230 GHz receivers to the 690 GHz receivers[6].